\documentclass{article}
\usepackage[utf8]{inputenc}
\usepackage{comment}

\title{Application of Homomorphic Encryption in Medical Imaging}
\author{
Francis Dutil, Alexandre See\footnote{Work done during an internship at Imagia} , Lisa Di Jorio, \\
and Florent Chandelier \\[1ex]
Imagia \\
\texttt{\small\{francis.dutil, alexandre.see, lisa.dijorio, florent.chandelier\}}\\ 
\texttt{\small@imagia.com}
}

\date{October 2021}

\usepackage[numbers]{natbib}
\usepackage{graphicx}
\usepackage{hyperref}
\usepackage[ruled,vlined]{algorithm2e}
\usepackage{subcaption}
\usepackage{multirow}
\usepackage{amsfonts}
\usepackage{stmaryrd}
\usepackage{amsmath}
\usepackage{amssymb}
\usepackage{caption}
\usepackage[font=footnotesize]{caption}

\begin{document}

\maketitle

\begin{abstract}
   In this technical report, we explore the use of homomorphic encryption (HE) in the context of training and predicting with deep learning (DL) models to deliver strict \textit{Privacy by Design} services, and to enforce a zero-trust model of data governance. First, we show how HE can be used to make predictions over medical images while preventing unauthorized secondary use of data, and detail our results on a disease classification task with OCT images. Then, we demonstrate that HE can be used to secure the training of DL models through federated learning, and report some experiments using 3D chest CT-Scans for a nodule detection task.
\end{abstract}

\section{Introduction}

Healthcare research is primordial for improving patient management \& care strategies, and greatly benefits society at multiple levels. However, most of this research requires access to large quantities of medical data, eventually including direct or indirect personal information. Privacy is of the utmost importance from both patient and ethic considerations perspectives. Personal health information should be accessed on a need-to-know basis in order to remain as confidential and secure as possible. Ideally, no health data should leave its fiduciary healthcare organization. Thus, research groups - either public, such as universities, or private, such as pharmaceutical companies - that wish to make use of such data need to design distributed processing strategies that mitigate the risk of exposing proprietary information (e.g. a proprietary machine learning model). 

Homomorphic Encryption (HE) is an emerging technology designed to process data that remains encrypted. Moreover, these operations produce encrypted results that can only be decrypted by the party holding the original encryption key. Using an appropriate design, HE can solve both the privacy \& governance challenges described above. In addition, some HE schemes are particularly appealing as they resist to most forms of attack, including quantum attacks.

Even though homomorphic encryption has shown promising results in multiple areas \cite{armknecht2015guide, bos2014private}, its use in combination with the latest artificial intelligence breakthroughs, such as deep learning based methods, remains an open challenge. The complexity of deep neural networks that can be hardened using homomorphic encryption is bounded by their heavy computational requirements. For example, the first proposed DL methods integrating HE took over 3 minutes to deliver a prediction over encrypted 28x28 monochrome thumbnail-like images (MNIST \cite{lecun1998gradient}); as a baseline, the equivalent plaintext inference takes a fraction of a second.

Although there exists a myriad of applications where homomorphic encryption would bring a lot of value, this paper focuses on two main applications: prediction, and federated learning.

In our prediction scenario, a hospital encrypts their medical data (image, CT scan, etc.) and sends it to a third party who performs DL inference and returns an encrypted prediction. For example, a hospital sends a homomorphically encrypted X-Ray to a cloud-based service to predict if the patient has chronic inflammatory lung disease (COPD). The cloud-based service then uses a proprietary algorithm to infer whether the patient has the disease, but its prediction remains obfuscated. By design, the hospital is the only one able to decrypt the HE prediction; accordingly, the inherent properties of HE protects the confidentiality of both the input patient data and the output diagnostic data, enabling ultimate protection against compromised third-parties, man-in-the-middle attacks and unauthorized secondary uses of data. As a complementary benefit, the cloud-based service stays in control of its intellectual property (i.e. the model). This scenario is also referred to as \textit{oblivious inference}.

In our federated learning scenario, multiple hospitals collaborate to train a model for a common task (such as predicting the presence of COPD, as introduced above), but do not share any patient data. This follows an established iterative methodology: (1) each hospital independently trains the same model locally on their own data, (2) the locally trained models are then sent to a unique third party server, (3) the third party server aggregates the weights over all models to finalize a globally trained model (an accepted baseline is weight averaging) and (4) the global model is sent back to each hospital to start another iteration. Hardening federated learning by using HE consists of (1) securing the communication between each client and the server, by encrypting the locally trained model prior to any transmission, and (2) constraining the server to use only HE-valid mathematical operations to perform model aggregation.  

For each of these scenarios, we study the feasibility and limitations of applying HE on real-world medical data. The rest of this paper is organized as follows: we provide a general overview of homomorphic encryption in Section \ref{sec:hereview}, before exploring inference in medical applications in Section \ref{sec:henn}. We detail how to use HE for secure aggregation in federated learning in Section \ref{sec:fedlearn}. Finally, we discuss and contextualize our results in Section \ref{sec:discussion}.

\section{Homomorphic Encryption Review}
\label{sec:hereview}

In this section, we present the general concepts of homomorphic encryption, and provide an overview of how to use them in the field of machine learning. 

Throughout this paper, we refer to \textbf{messages} as $m_1, m_2, m_3, ...$. They correspond to values in clear text (plaintexts). They can be single values, vectors, matrices, depending on context. \textbf{ciphertexts} $c_1, c_2, c_3, ...$ refer to encrypted messages. In the case of Ring Learning With Rounding (RLWR, covered in Section \ref{subsec:ringlearning}), a ciphertext is represented in a lattice. 

Formally, an encryption scheme is homomorphic if, for a given encoding function $Enc$, and decoding function $Dec$, there exists operations $\{\diamond,\circ\}$ such that

\begin{equation}
\begin{split}
Dec(Enc(m_1) \diamond Enc(m_2)) = m_1 \circ m_2,
\end{split}
\end{equation}

where typically $\{\diamond,\circ\}$ are multiplication, addition, etc. We can thus have schemes of different strength depending on the $Dec, Enc$ function. 

More broadly, there are three types of homomorphic encryption:

\begin{itemize}
    \item \textbf{Somewhat homomorphic encryption} (SHE) refers to schemes that only support a subset of mathematical operations. For example, the Paillier \citep{paillier1999public} scheme only supports the addition of two ciphertexts $c_1 + c_2$.
    \item \textbf{Leveled homomorphic encryption} (LHE) refers to schemes that support $\{+, \times\}$ operations, but only a limited number of times (levels) \citep{brakerski2012LHE}. For example, $c_1 \times c_2 \times c_3$ has two levels of multiplication. The level of a particular implementation is usually determined by the number of multiplications, which is typically the most expensive operation.
    \item \textbf{Fully homomorphic encryption} (FHE) refers to schemes that can apply $\{+, \times\}$ operations an arbitrary number of times. \citep{gentry2009fully} was the first to come up with a FHE scheme.
\end{itemize}

It is important to note that FHE schemes, even though more "powerful" on the encryption strength, should not always be used as they require extensive and sometimes unaffordable compute time; in the first implementation of \citep{gentry2009fully}, one (1) multiplication took around 30 minutes to complete.

Each type of homomorphic encryption has their own ideal context of use: SHE can be used in federated learning since no multiplications are involved in the aggregation (see Section \ref{sec:fedlearn}); LHE schemes are great when working with small networks that don't have a big multiplicative depth; and FHE can be used for training on encrypted data. 

In the next sections, we describe two important homomorphic encryption schemes for our use cases: Ring Learning With Error and Paillier's.

\subsection{Ring Learning With Error}
\label{subsec:ringlearning}

This paper is not meant to be an in-depth tutorial on the theory of homomorphic encryption, the interested reader should refer to blog posts \footnote{\href{https://medium.com/asecuritysite-when-bob-met-alice/learning-with-errors-and-ring-learning-with-errors-23516a502406}{Learning With Errors and Ring Learning With Errors}} \footnote{\href{https://blog.openmined.org/build-an-homomorphic-encryption-scheme-from-scratch-with-python/}{Build an homomorphic encryption scheme from scratch with python}}, and this review \citep{aslett2015review}. Nonetheless, here are a few takeaways without diving too much into the involved mathematics. 

Let $s$ be the secret key, and $(A,b)$ the public key created by sampling $A\sim \epsilon_1$ and generating $b = A \times s + \epsilon_2$, with $\epsilon$ coming from a random distribution. Let $m \in \mathbb{Z}$ be an integer encoded in a vector using its binary representation.

Given $s$, $A$, and $b$, the $Enc$ function can be defined as follows:

\begin{equation}
\begin{split}
Enc(m) = (c_1, c_2) = (A + \epsilon_3, b + m + \epsilon_4),
\end{split}
\end{equation}

and the $Dec$ function as follows (remove $A \times s \approx b$ to recover $m$)

\begin{equation}
\begin{split}
Dec(c_1, c_2) = m_{d} \approx c_2 - c_1 \times s
\end{split}
\end{equation}

Notice that the decrypted message is only approximate, because of the noise $\epsilon$ that was added throughout the scheme. The trick here is that \textbf{if the noise is small enough}, we can round the decrypted $m$ to the closest integer.

If we have another message $m^\prime$ encoded in $(c_3, c_4)$ we can add $c_2 + c_4$ without changing anything. But multiplying them to multiply messages also multiplies messages and noises.

The intuition behind the scheme's security is to hide the message in the forest (noise). The difficulty of this encryption scheme lies in managing this noise. It needs to be small enough so that we can decrypt the message, but big enough so that the scheme stays secure. We also need to take into consideration the fact that the noise increases each time we do some calculation on the ciphertexts. This is particularly true for multiplications, which is why, throughout this paper and in the literature, people will often refer to the \emph{multiplicative depth} of a computation, which represents the number of consecutive multiplications we can perform on a ciphertext before the noise corrupts the underlying message.

A typical way of increasing the multiplicative depth limit is to increase the size of the keys. Of course, this comes at a computational cost.

A way to increase the schema efficiency while keeping the same level of security is to encode ciphertexts in polynomial Rings (hence the name \emph{Ring Learning With Error}). This version substitutes the space $\mathbb{Z}_n$ (integers in $\llbracket0, n-1\rrbracket$) by using the polynomial quotient ring $\mathbb{Z}_n / (x^p+1)$. In this space, elements are encoded into polynomials of degree lower than $p-1$. Indeed, the $(x^p+1)$ stands for the coefficient in modulo, (for example $x^4 \equiv -1 \pmod{(x^4+1)}$), such that degrees are no higher than $x^{p-1}$. In doing so, we enable a higher complexity and better security of the encoding, as well as the possibility to encode vectors through their polynomial representation.

\textbf{CKKS encoding:} The way in which plaintexts are encoded in ciphertexts is of a particular importance. For example, encoding a single message in a ciphertext would be extremely wasteful, as we deal with millions, if not billions of numbers. An encoding scheme that is particularly well suited for this kind of situation is CKKS \cite{cheon2017homomorphic}. It allows us to not only encode multiple thousands of floating-point values in the same ciphertext, but also to rotate values inside it without burning any of our multiplicative budget. This feature will be extremely important later on, since it provides an efficient way to compute matrix multiplication.

\subsection{Paillier's encryption scheme}
\label{subsec:paillier}

Another prominent scheme used is Paillier's encryption \cite{paillier1999public}, where the hardness of the problem relies on the intractability of computing discrete logarithms. However, knowledge of the secret key allows for the efficient decryption of ciphertexts.

Briefly, here are the most important aspect of the algorithm:

\textbf{Key generation:} Let $n = pq$ where $p,q$ are prime of equivalent length. Let $g$ be an integer of order $n\alpha \textbf{ mod } n^2$, with $\alpha \geq 1$. The public key is $(n, g)$, and the secret key is $\lambda (n)$ (denotes the Carmechael's lambda function), which is defined as the largest order of the element of $\mathbb{Z}_n^*$. Simply put, $\lambda(n)$ is the least common multiple of $(p-1)$ and $(q-1)$. 

\textbf{Encryption:} To encrypt a message $m \in \mathbb{Z}_n$, we randomly sample $x \in \mathbb{Z}_n^*$, then compute the ciphertext

\begin{equation}
\begin{split}
Enc(m) = c = g^m x^n \textbf{ mod } n^2
\end{split}
\end{equation}

Unlike RLWE described earlier, this scheme only supports the addition of ciphertext, not multiplication. More formally, the scheme has the following properties:

\begin{equation}
\begin{split}
E(m_1 + m_2) = E(m_1) \times E(m_2)\\
E(k \times m) = E(m)^k \textrm{, with k }\in \mathbb{Z}
\end{split}
\end{equation}

\textbf{Decryption:} To decrypt a cyphertext $c$, we compute

\begin{equation}
\begin{split}
m = \frac{L(c^{\lambda(n)} \textbf{ mod } n^2)}{L(g^{\lambda(n)} \textbf{ mod } n^2)},
\end{split}
\end{equation}

where $L(u) = \frac{u-1}{n}$. The Taylor expansion of the logarithm allows understanding how this relates to discrete logarithm :
\begin{equation}
\begin{split}
(1+n)^x & = 1+n x+{2 \choose x} n^2 + ... \\
& \equiv 1+n x \pmod{n^2}
\end{split}
\end{equation}

\begin{equation}
\textrm{so, } x \equiv \frac{y-1}{n} \pmod{n^2} \textrm{, with } y := (1+n)^x \pmod{n^2}
\end{equation}

\begin{equation}
\textrm{so finally, } x \equiv L(y) \equiv L((1+n)^x) \pmod{n^2}
\end{equation}

The last equation shows that $L$ can stand as an approximation of the logarithm. To get more details about the scheme, as well as a proof of its security, we refer the reader to the original paper.

\subsection{Paillier's threshold variant}
\label{sec:ttp}

A variant of Paillier's encryption scheme allows for the decentralization of the decryption process. This requires that the private key be fragmented into \textit{secret shares} and shared between clients. This design allows to simply exclude a compromised client from any future round. It also has the added benefit of allowing the decryption of models even if not all clients have participated (in case of connection dropout, for example).

There exists multiple ways to render Paillier's scheme "decentralized", and we describe in the following section the one from \cite{fouque2000sharing}.

Shamir's secret sharing scheme \cite{shamir1979share} is at the core of this approach, and is based on the Lagrange interpolation theorem. This theorem states that for any polynomial of degree $l-1$, only $l$ samples of said polynomial are necessary in order to reconstruct it (i.e. given $l$ points, there's only one polynomial of degree $l-1$ that connects them). More formally, for any polynomial $P$, we have: 

\begin{equation}
\begin{split}
P(x) = \sum_{i=1}^l \prod_{j\neq i}^l \frac{x-x_j}{x_i-x_j} P(x_i)
\end{split}
\end{equation}

In a federated learning application, $l$ represents the number of clients.

The secret shares are thus generated by first sampling a random polynomial of degree $l-1$, $P(x) = c_0 + c_1 x + ... + c_{l-1}x^{l-1}$. Each of the $l$ clients is then given a single observation of a non-zero integer value: $sk_i = P(i), i \neq 0$. 

Given all of those keys, the polynomial can be reconstructed, and the secret key recovered. The secret key is typically $P(0)$:

\begin{equation}
\begin{split}
sk = P(0) = \sum_{i=1}^l \prod_{j = 1, j\neq i}^l \frac{x_j}{x_j - x_i} sk_i
\end{split}
\end{equation}

The trick is then to find a way to integrate this Lagrange interpolation in our encryption scheme. Formally, Shamir's secret sharing is incorporated in Paillier's encryption scheme as follows:

\textbf{Key generation:} Choose an integer $n = pq$, where $p,q$ are strong prime, i.e. $p = 2p' + 1, q = 2q' + 1$, with $p', q'$ primes. Let $n' = p'q'$. Let $\beta$ be randomly picked from $\mathbb{Z}_n^{*}$. Randomly select $(a,b) \in \mathbb{Z}_n^{*} \times \mathbb{Z}_n^{*}$. Set $g = (1+n)^a b^n \textbf{ mod } n^2$. The public key is $\theta = L(g^{\beta n'})$. The secret key $\beta n'$ is shared using the Shamir's scheme. As discussed above, we create a random polynomial of degree $l-1$, where the constant term is our secret key: $P = \beta n' + c_1 x + ... c_{l-1} x^{l-1}$. The secret key of each client $i$ is then $sk_i = P(i) \textbf{ mod } n n'$.

\textbf{Encryption:} As in the original scheme, to encrypt a message $m \in \mathbb{Z}_n$, we randomly sample $x \in \mathbb{Z}_n^*$, then compute the ciphertext

\begin{equation}
\begin{split}
c = g^m x^n \textbf{ mod } n^2.
\end{split}
\end{equation}

\textbf{Decryption:} Unlike in the original scheme, no entity can decrypt a ciphertext on their own. To do so, a round of communication between all keyholder is necessary. First, each $i^{th}$ client computes it's partial decryption

\begin{equation}
\begin{split}
c_i = c^{2\Delta sk_i} \textbf{ mod }n^2,
\end{split}
\end{equation}

where $\Delta = l!$. After a round of communication where each client has received everyone else's partial decryption, we can finally fully decrypt $c$:

\begin{equation}
\begin{split}
m = L(\prod_{j=1}^l c_j^{2\mu_{j}} \textbf{ mod } n^2 ) \times \frac{1}{4\Delta^2 \theta} \textbf{ mod } n
\end{split}
\end{equation}
, where $\mu_j = \Delta \prod_{j'=1, j' \neq j}^l \frac{j'}{j'-j} \in \mathbb{Z}$.

\section{Using Homomorphic Encryption with Neural Networks}
\label{sec:henn}

Work combining homomorphic encryption with neural network are numerous, and the domain is in constant evolution.

For oblivious inference, the main line of work in homomorphic encryption applied to machine learning comes from CryptoNets \citep{gilad2016cryptonets}. \citep{brutzkus2019low} proposes further improvements to their network. For example, a prediction on MNIST takes less than a second, and a prediction on cifar10 with a 2 layer CNN can be performed in approximately 10 minutes. For these papers, only the leveled version of RLWE is used, since the bootstrapping operation would be computationally prohibitive. 

Other approaches like GAZELLE \cite{juvekar2018gazelle}, XONN \cite{riazi2019xonn} and MiniONN \cite{liu2017oblivious} are built on top of these techniques by incorporating some MPC elements (like garble circuits) to various degrees. Although these techniques have the advantage of allowing the computation of non-polynomial functions (like ReLU), they add a communication burden. 

For federated learning, the FATE \cite{zhang2020batchcrypt} algorithm uses Paillier's encryption scheme. However, all clients share the same secret key, which is highly problematic if one of them is compromised. Another approach can be found in \cite{truex2019hybrid}, where they use a threshold variant of Paillier's scheme in addition to differential privacy. The used threshold variant guaranties the security of the experiment even if a client is compromised.

\section{Predicting over Encrypted Medical Data using Oblivious Inference}
\label{sec:obliviousinf}

One of the biggest challenges in applying homomorphic encryption to deep learning is finding a way to compute matrix multiplication efficiently. 

A trivial way of doing so would be to encode each element of the matrix in its own ciphertext, and then compute the result as usual. However, this doesn't leverage the encoding power of some encryption schemes like CKKS. Using such techniques, one can encode entire vectors (or a whole matrix) into a single ciphertext. This effectively reduces the computation time of a multiplication by the number of elements in this vector.

A small caveat with this encoding technique is that we can't use the traditional way of computing matrix multiplication. Instead, we need to use the diagonal order matrix multiplication\citep{halevi2014algorithms} technique. Briefly, for a non-encrypted matrix $A$ and encrypted vector $b$, $A \times b$ is computed by first arranging $A$ in its diagonal form, and then applying a series of rotation and multiplication. For a matrix $A$, its diagonal order $A^{diag}$, and a vector x, we can compute $A \times x = b$ by doing 

\begin{equation}
\begin{split}
b = \sum_i A^{diag}_i \odot rotation(x, i),
\end{split}
\end{equation}

where the rotation shifts all the elements to the left. Since a rotation in the polynomial doesn't use any multiplicative budget, we thus only use a single multiplicative level to perform a matrix multiplication.

\begin{figure}
\begin{subfigure}{.33\textwidth}
  \centering
  \includegraphics[width=.8\linewidth]{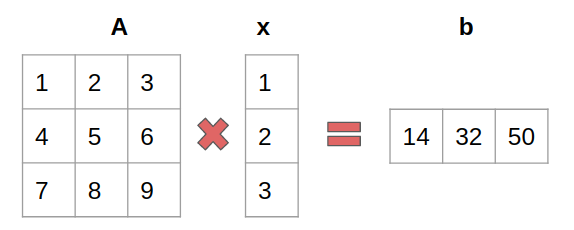}
  \caption{}
  \label{fig:normalmatmul}
\end{subfigure}%
\begin{subfigure}{.33\textwidth}
  \centering
  \includegraphics[width=.8\linewidth]{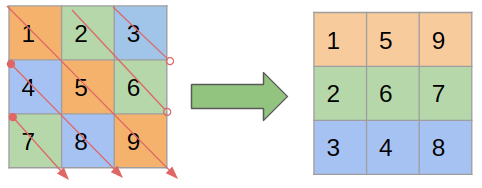}
  \caption{}
  \label{fig:diagorder}
\end{subfigure}
\begin{subfigure}{.33\textwidth}
  \centering
  \includegraphics[width=.8\linewidth]{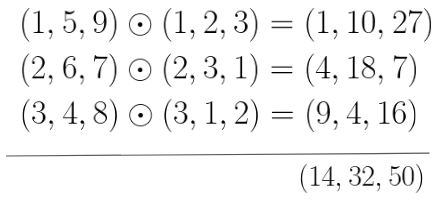}
  \caption{}
  \label{fig:diagmatmul}
\end{subfigure}
\caption{Diagonal order matrix multiplication. In (\ref{fig:normalmatmul}) we have a typical matrix multiplication of $A\times x = b$. (\ref{fig:diagorder}) shows how to get the diagonal order of a matrix, and (\ref{fig:diagmatmul}) how to compute the matrix multiplication with a series of rotation and element wise multiplication.}
\label{fig:matmul}
\end{figure}

A visualization of the technique can be found in figure \ref{fig:matmul}, and more details for our specific case in algorithm \ref{alg:matmul}. 

\begin{algorithm}[H]
\SetAlgoLined

\SetKwInOut{Input}{input}
\Input{Matrix A // Non encrypted matrix}
\Input{ciphertext b // Encrypted vector}
\KwResult{c, result of A$\times$b}
$A_{diag} = Diag(A)$// See Algorithm \ref{alg:diag}\;
 c = empty ciphertext \\
 b = b + Rotation(b, -cols(A$_{diag}$)) // Simulate a $mod$ $rows(A)$ in the encrypted space.\\
\For{$i\gets0$ \KwTo $rows(A_{diag})$}{
    $row_i = A_{diag}[i]$ \\
    $t_i$ = Rotation(b, i) // Rotate the ciphertext. \\
    $t_i = row_i \odot b_i$ // Element wise multiplication in the encrypted space. \\
    $c = c + t_i$ // Element wise addition in the encrypted space. 
    }   
 
 return c
 \caption{Matrix-vector multiplication}
 \label{alg:matmul}
\end{algorithm}

\begin{algorithm}[H]
\SetAlgoLined
\SetKwInOut{Input}{input}
\Input{Matrix A // Non encrypted matrix}
\KwResult{$A_{diag}$, matrix in diagonal order}
\uIf{A is not square}
{pad A with 0s}
 $A_{diag} = zerosLike(A)$ \\

   \For{$i\gets0$ \KwTo $rows(A)$}{
       \For{$j\gets0$ \KwTo $cols(A)$}{
        $A_{diag}[i,j] = A[j, j+i]$ // All indexes are $modulus$ the array size.  
        }   
    }

 return $A_{diag}$
 
 \caption{Matrix diagonalization}
 \label{alg:diag}
\end{algorithm}

\subsection{2D convolution}

\begin{figure}
\begin{subfigure}{.5\textwidth}
  \centering
  \includegraphics[width=.8\linewidth]{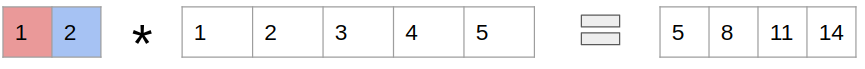}
  \caption{}
  \label{fig:conv1d}
\end{subfigure}%
\begin{subfigure}{.5\textwidth}
  \centering
  \includegraphics[width=.8\linewidth]{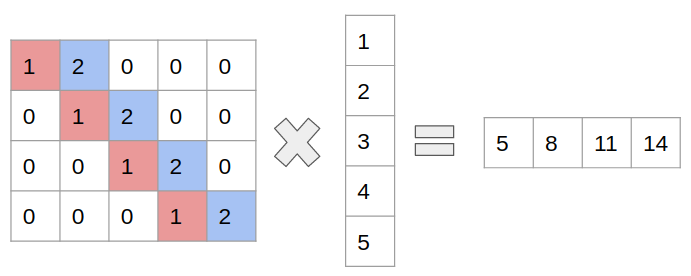}
  \caption{}
  \label{fig:convtoeplitz1d}
\end{subfigure}
\caption{Viewing a convolution as a matrix multiplication. In (\ref{fig:conv1d}) we have a typical convolution, with filter = $(1,2)$. In (\ref{fig:convtoeplitz1d}) we see the same convolution, but seen as a matrix multiplication. The filter $(1,2)$ turned into a Toeplitz matrix.}
\label{fig:toeplitz1d}
\end{figure}

A straightforward way to come up with an algorithm for convolution is to simply view the whole process as a single matrix multiplication operation. Indeed, the weight matrix could be transformed into a Toeplitz matrix 
\footnote{Please check out the wikipedia entry on Toeplitz matrices:  \href{https://en.wikipedia.org/wiki/Toeplitz_matrix}{https://en.wikipedia.org/wiki/Toeplitz\_matrix}}, then the algorithm \ref{alg:matmul} can be used. An example for 1d convolution can be found in figure \ref{fig:toeplitz1d}. This technique \emph{could} rather elegantly be extended to 2D and 3D convolution, but scaling to real-world problems quickly become impossible.  

The sheer size of the resulting vector prevents from generalizing to 2D and 3D convolution: unlike a dense layer which typically have at most 1000 elements, convolutions can quickly jump to tens of thousands, or even hundreds of thousands of elements, since every feature maps needs to be encoded.

For example, a small image of size $28\times28\times64$ contains 677376 elements. However, in SEAL a ciphertext encoded with CKKS and 32768 security bytes (the maximum possible value) can hold at most 16384 elements. We clearly need to encode all those values in different ciphertexts.

In \cite{juvekar2018gazelle}, each feature map is encoded in a different ciphertext. Features of size $28\times28\times64$ would thus be encoded in 64 different ciphertexts. This works well for small images, but images bigger than $\sim$100x100 pixels cannot be processed.

In order to process bigger images, which are very common in medical imaging, we instead encode the images row by row. Features of size  $28\times28\times64$ would thus be encoded in 28 ciphertexts.

We then need to process these ciphertexts by a series of 1 dimensional convolutions.

Algorithm \ref{alg:conv2d} describes how to perform 2D convolution using multiple 1D convolution.

\begin{algorithm}[h]
\SetAlgoLined
\KwResult{2D convolution: A$*$B }
\SetKwInOut{Input}{input}
\Input{Filters A // Non encrypted convolution parameters. 4D array: (height, width, input channel, output channel).}
\Input{ciphertexts B //  List of encrypted vectors. Each encoding a specific row}

output\_height = len(B) - A.height + 1 // image output height \\
output\_width = len(B) - A.width + 1 // image output width \\

ciphertexts res // list of encrypted rows (output)

\For{$i\gets0$ \KwTo output\_height}{
\For{$j\gets0$ \KwTo A.height}{

    // For this filter$_j\times$B$_i$, compute all output channels. \\
    // Add it to the corresponding cipher text. \\
    \For{$k\gets0$ \KwTo A.output\_channel}{
    
        sub\_A = A[j, :, :, k]  \\
        sub\_A = Toeplitz(sub\_A) \\
        
        c$_k$ = matmul(sub\_A, B$_{i+j}$) // Algorithm \ref{alg:matmul} \\
        c$_k$ = Rotation(c$_k$, -output\_width$\times$k)\\
        res$_i$ = res$_i$ + c$_k$
        }   
    
    }   
}   

\caption{2D convolution}
\label{alg:conv2d}
\end{algorithm}

\subsection{Global Average Pooling}

Most modern convolution neural networks now have a \emph{Global Average Pooling} layer. This layer aggregates the features across all spatial position. For example, if an image is of size $28\times 28 \times 64$, this produces a vector of size 64.

The full procedure can be found in algorithm \ref{alg:gap}. A small but important detail is that the resulting ciphertext also contains some garbage values in between the proper values. It's impossible to remove those unwanted values without masking, which would unnecessarily burn our multiplicative budget. Instead, we keep track of this condition until the next time a matrix multiplication is done. This way, we can apply the masking on the clear weights, saving us a multiplicative level.

The same strategy can be applied to the \emph{averaging} part. Instead of multiplying the resulting ciphertext by $H_i\times W_i$, we can keep track of this value and scale the next weights by it instead.

\begin{algorithm}[h]
\SetAlgoLined
\KwResult{Something }
\SetKwInOut{Input}{input}
\Input{ciphertexts A //  List of encrypted vectors. Each encoding a specific row}
\Input{width // The number of column in the image.}

ciphertext sum$\_$row \\
ciphertext result \\
// Sum over all the rows. \\
\For{$i\gets0$ \KwTo len(A)}{
    sum$\_$row = sum$\_$row + A[i]
}   

// Sum over all the columns. \\
\For{$i\gets0$ \KwTo width}{
    res = res + rotation(sum$\_$row, i)
}   

 \caption{Global average pooling}
 \label{alg:gap}
\end{algorithm}

\subsection{Experiments and Results}

In this section, we show how to use oblivious inference with a type of real-world medical data, Optical Coherence Tomography (OCT) images. OCT is a non-invasive procedure, in which light waves are used to take cross-section pictures of the retina. Those images map and measure the thickness of the retina, and help the ophthalmologist diagnose a variety of diseases.

\begin{figure}
\begin{subfigure}{.5\textwidth}
  \centering
  \includegraphics[width=.8\linewidth]{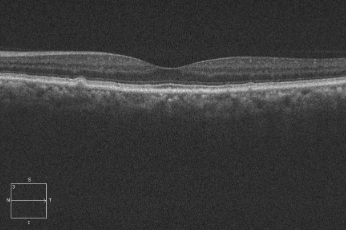}
  \caption{}
  \label{fig:octid_amd}
\end{subfigure}
\begin{subfigure}{.5\textwidth}
  \centering
  \includegraphics[width=.8\linewidth]{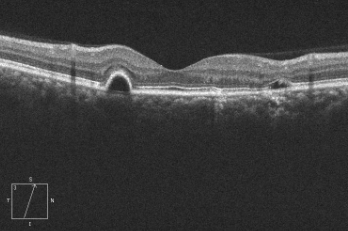}
  \caption{}
  \label{fig:octid_csr}
\end{subfigure}
\begin{subfigure}{.5\textwidth}
  \centering
  \includegraphics[width=.8\linewidth]{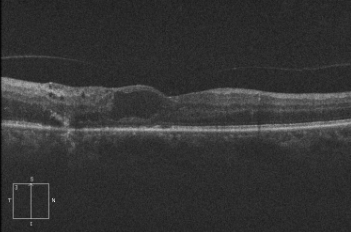}
  \caption{}
  \label{fig:octid_dr}
\end{subfigure}
\begin{subfigure}{.5\textwidth}
  \centering
  \includegraphics[width=.8\linewidth]{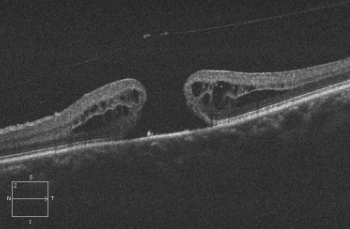}
  \caption{}
  \label{fig:octid_mh}
\end{subfigure}
\begin{subfigure}{.5\textwidth}
  \centering
  \includegraphics[width=.8\linewidth]{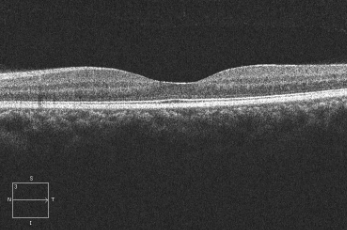}
  \caption{}
  \label{fig:octid_normal}
\end{subfigure}
\caption{Viewing the five classes of the OCTID dataset. \ref{fig:octid_amd} is AMD, \ref{fig:octid_csr} is CSR, \ref{fig:octid_dr} is DR, \ref{fig:octid_mh} is MH, and \ref{fig:octid_normal} is normal}.
\label{fig:octid_dataset}
\end{figure}

We used the OCTID \cite{gholami2020octid} public dataset, which consists of 572 images. All scans were captured with a Cirrus HD-OCT machine at Sankana Nethralaya Eye Hospital, in India. The dataset consists of normal retinas and samples from 4 different diseases, Macular Hole (MH), Age-related Macular Degeneration (AMD), Central Serous Retinopathy (CSR), and Diabetic Retinopathy (DR). Samples of each of the different classes can be found in figure \ref{fig:octid_dataset}.

For this experiment, all images were resized from 500x750 to 112x112 pixels. The dataset is split 60\% for the train, 10\% for the valid, and 30\% for the test set. The different sets are stratified by pathology to guaranty a good distribution. 

The computational constraints imposed by homomorphic encryption only allow for shallow networks. For this reason, we use a Convolutional Neural Network (CNN) with 2 or 3 convolution layers. Each layer has a kernel size of 5x5 with strides of 4. 
Additionally, since only polynomials can be evaluated homomorphically, the square function $f(x) : x^2$ is used as non-linearity instead of the typical ReLU \cite{gilad2016cryptonets}.

Given the low number of samples, the network is first pretrained on Imagenet \cite{deng2009imagenet}, and then subsequently on Mendeley \cite{kermany2018large}, another OCT dataset.

\begin{figure}
\centering
\includegraphics[width=0.9\linewidth]{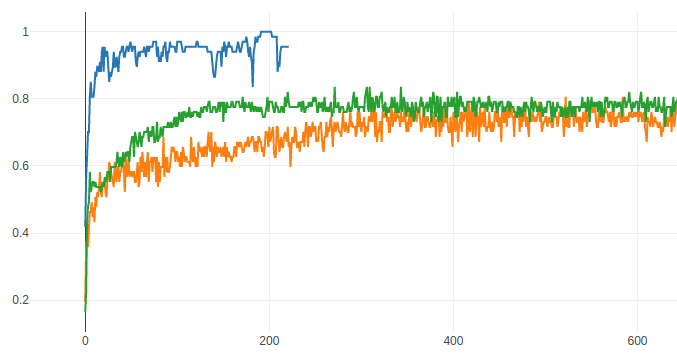}
\caption{Validation accuracy of different networks on the OCTID dataset. In blue a VGG16, in green a 3 layer CNN, and in orange a 2 layer CNN.}
\label{fig:inf_curve}
\end{figure}

\begin{table}[]
\begin{tabular}{|l|c|c|}
\hline
Neural network & Test accuracy & \begin{tabular}[c]{@{}c@{}}Average inference time \\ (minutes)\end{tabular} \\ \hline
VGG16          & 95\%          & -                                                                           \\ \hline
2 layer CNN    & 82\%          & 22                                                                          \\ \hline
3 layer CNN    & 79\%          & 180                                                                         \\ \hline
\end{tabular}
\caption{Test accuracy, and inference time when computed on encrypted data. The time corresponds to the average time to process a single example. Inference was done on two Intel Xeon Processor E5-2650 v4, for a total of 24 cores.}
\label{tab:inf}
\end{table}

The learning curves of a VGG16, a 2 layer CNN and a 3 layer CNN can be found in figure \ref{fig:inf_curve}. Clearly, with their large number of parameters, typical neural networks are better suited for this type of task. However, smaller neural network are still able to achieve a reasonable performance. 

The inference time when processing encrypted data can be found in table \ref{tab:inf}, along with the test score. As we can see, adding a single layer increases drastically the time it takes to process one example, from 20 minutes to 3 hours. 

\section{Secure Federated Learning using Paillier's \\Encryption Scheme}
\label{sec:fedlearn}

\begin{figure}
    \centering
    \includegraphics[scale=0.5]{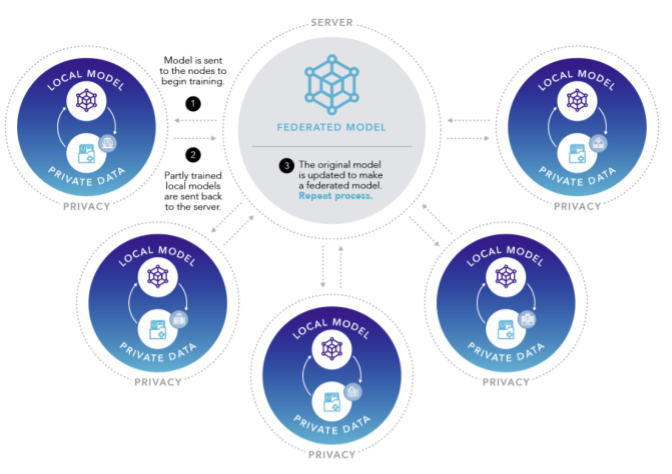}
    \caption{General overview of Federated Learning. Multiple institutions, shown in blue, collaborate to learn a common model using a third-party server (in gray) in order to securely aggregate their learning.}
    \label{fig:fl_general}
\end{figure}

A second domain of application for homomorphic encryption in deep learning is federated learning. A comprehensive overview can be found in Figure \ref{fig:fl_general}, and an algorithmic representation can be found in algorithm \ref{alg:federatedlearning}.

\begin{algorithm}[h]
\SetAlgoLined
\SetKwInOut{Input}{input}
Server executes:\\
\hspace*{10pt}initialize $\omega_0$\\
\hspace*{10pt}\For{each round t = 1, 2,...}{
\hspace*{10pt}    $m \leftarrow \text{max}(C . K, 1)$\\
\hspace*{10pt}    $S_t \leftarrow \text{(random set of m clients)}$\\
 \hspace*{10pt}   \For{each client $k \in S_t$ in parallel do}{
\hspace*{10pt}        $\omega_{t+1}^{k} \leftarrow \text{ClientUpdate}(k, w_t)$ \\
 \hspace*{10pt}       $\omega_{t+1} \leftarrow \frac{1}{K}\sum_{k=1}^{K} \omega_{t+1}^{k}$
    }
}

\textbf{ClientUpdate}$(k, \omega)$: // Run on client k \\
\hspace*{10pt}\For{each local batch b = 1,2,... $\mathcal{B}$}{
\hspace*{10pt} 
$\omega \leftarrow \omega - \eta \triangledown \ell (\omega ; b)$
}
return $\omega$ to server
 \caption{Traditional federated learning. The $K$ clients are indexed by $k$; $B$ is the local number of minibatch on a round, $E$ is the number of local epochs, and $\eta$ is the learning rate.}
 \label{alg:federatedlearning}
\end{algorithm}

In this scenario, security relies on Paillier's scheme (as explained in Section \ref{subsec:paillier}), which is implemented during encrypted client networks aggregation. We use weight averaging to aggregate the client networks. Since Paillier's scheme is only \textit{somewhat homomorphic}, the final averaging division is done locally on each client after decryption.

Once the aggregation is done, the clients collaborate to decrypt the resulting network, as explained in \ref{sec:ttp}. Since hundreds of thousand of weights are encoded, the main challenge revolves around efficiently encoding networks.

An efficient yet simple approach is to encode multiple weights in a single integer. This is possible because the modulo ring in which we work is gigantic (thousands of bits), and with enough bits of padding between each weight, we can avoid overflow during addition. Negative value are taken care of by first shifting all values by the same constant.

\subsection{Experiments and Results}

We used the Lung Image Database Consortium (LIDC) \cite{armato2011lung}, a well-characterized repository for chest computed tomography (CT). The LIDC/IDRI database contains 1018 cases from seven academic centers and eight medical imaging companies. The images are annotated by four experts; in total the database contains 7371 lesions marked “nodule” by at least one radiologist, including 2669 nodules with a size greater than three millimeters. Among these 2669 nodules, 928 were annotated at the same location from all 4 experts. 

For this specific set of experiments, only nodules greater than 4 millimeters are considered. Volumes are converted to Hounsfield units and voxel sizes are 1.25 millimeters isotropic 48 x 48 x 48 voxels in the subvolume. Subvolumes are created to ensure that the nodule appears at random locations to prevent an algorithm from learning the nodule's position.

\begin{figure}
\centering
\includegraphics[width=1.\linewidth]{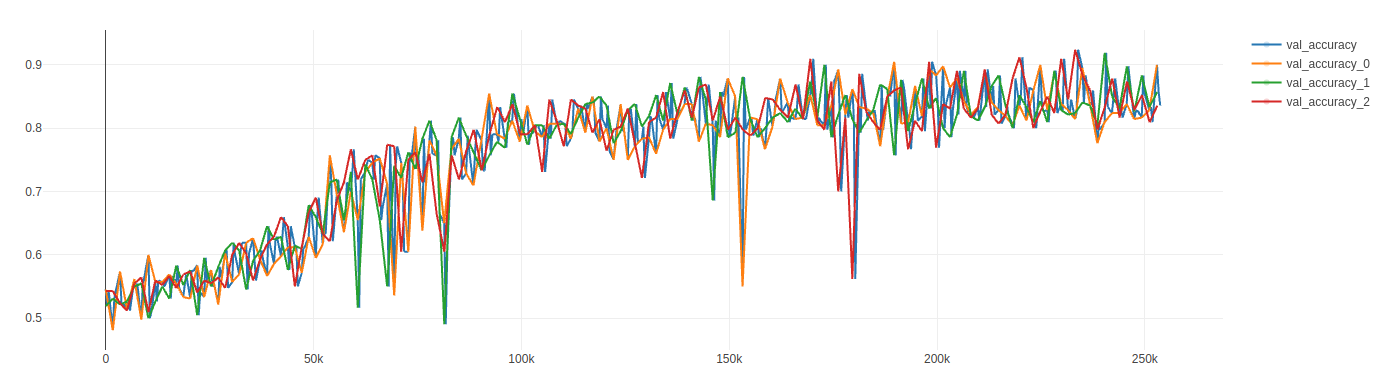}
\caption{Federated learning using the threshold variant of Paillier's encryption scheme. The validation accuracy of the three clients are $val\_accuracy\_0, val\_accuracy\_1, val\_accuracy\_2$, and the accuracy with the aggregated network is $val\_accuracy$.}
\label{fig:ttp_curve}
\end{figure}

The network is a 3D variant of the squeezenet network \cite{iandola2016squeezenet}. The training set is split into 3 distinct subsets to simulate 3 clients. For the encrypted part, a key of 2048 bits is used. The learning curve can be found in \ref{fig:ttp_curve}.

\textbf{Task performance:} Both the encrypted and control pipelines lead to 90\% accuracy on the validation set.

\textbf{Clock time performance:} On average, it takes around 13 minutes to complete a full round. 11 minutes to train the model, 1 minute to fully encrypt the model's 400k parameters, 1 second to aggregate the weights of the different clients, then 1 minute for each client to decrypt the model.

\textbf{Setup:} All experiments is done on a single machine with excess available RAM (755 GB). Multiprocessing is use to accelerate the encryption and decryption operations using two Intel Xeon E5-2698 v4 2.2 GHz processors, with 60 of the 80 available threads. Each client is trained using its own NVIDIA Corporation GV100GL GPU [Tesla V100 PCIe 32GB] (rev a1). 

\section{Discussion}
\label{sec:discussion}

In this study, we have investigated the use of homomorphic encryption (HE) concurrently with deep learning methods, to demonstrate a robust privacy by-design solution to applying AI in the healthcare context. Our results suggest that using HE for oblivious inference is appropriate for clinical workflows that are not time sensitive. An interesting observation is that adding a single layer to a deep learning architecture leads to an exponential increase in computation time. Indeed, using leveled HE renders the lower bound of a ciphertext's size proportional to the depth of the network. 

However, we believe that there is room for improvement. First, multiprocessing (e.g.: over GPUs) offers a natural way to reduce the computational cost of HE operations, but existing machine learning frameworks remain poorly adapted. Second, while efficient bootstrapping implementations remain elusive, although research in this area is very active, they could be exploited to limit the effective multiplicative depth of arbitrary architectures. Finally, another promising avenue is the creation of minimal architectures highly optimized for specific tasks over specific imaging modalities.\\

We further evaluated the effectiveness of HE to establish a 0-trust security model, while protecting both the confidentiality \& the privacy of information exchanged during a federated learning experiment. We practically demonstrated that our solution can be used, within the same compute time \& AI performances assumptions, to offer training services to organizations that cannot afford to share, transfer or disclose information between each others. The proposed method addresses the short-comings related to current differential privacy techniques traditionally used to limit privacy attacks but often at the price of decreasing AI performances \citep{Kaissis2021}. Finally, our implementation maintains all principles of traditional federated learning methods, and accordingly it can benefit from all research innovations in the field such as split learning \cite{vepakomma2019}.

\section{Conclusion}

In this study, we provided practical solutions to address challenging privacy concerns limiting the use of AI in healthcare, by applying homomorphic encryption (HE) on top of deep learning methods. More specifically, we demonstrated the effectiveness of HE, in the context of  a clinical decision support system for the diagnostic of diabetic retinopathy, to predict over encrypted real-world medical images. We further explored the use of HE to facilitate collaborative research across distributed datasets within a federated learning framework, and illustrated how both privacy \& trust requirements can be delivered. We stress-tested this solution by training a computer-aided detection system with application to lung cancer screening from 3D CT scans, and reported promising performances with potential to generalize across similar deep learning detection \& diagnostic tasks.

We believe that current advances in homomorphic encryption are a path forward to facilitate the use of AI in healthcare. The complexity of the math involved comes with the great benefits of enabling quantum-resistant privacy \& confidential protection, 0-trust security models, and strong governance over secondary uses of data. Our future directions include the exploration of practically training deep learning models directly over encrypted data under reasonable compute time assumptions.

\bibliographystyle{plain}
\bibliography{references}
\end{document}